\documentclass[MSNbibl,nameyear,dvips]{arxstspdf}
\usepackage{dcolumn}
\usepackage{flushend}
\usepackage{stfloats}
\usepackage{url,breakurl}
\volume{29}
\issue{2}
\pubyear{2014}
\firstpage{181}
\lastpage{200}
\doi{10.1214/13-STS462} 

\makeatletter
\newcolumntype{d}[1]{D{.}{.}{#1}}
\newproclaim{Note}{Note}
\newcommand{\Rexpr}[1]{\texttt{#1}}
\newcommand{\Rfunc}[1]{\textit{#1()}}
\newcommand{\Rpkg}[1]{\textbf{#1}}
\newcommand{\Rvar}[1]{\texttt{#1}}
\newcommand{\Rkeyword}[1]{\textsl{#1}}
\newcommand{\OmgPkg}[1]{\textbf{\textit{#1}}}
\newcommand{\BiocPkg}[1]{\textbf{\textsl{#1}}}
\newcommand{\Cpp}{\textsl{C}$++$}
\newcommand{\C}{\textsl{C}}
\newcommand{\R}{\textsl{R}}
\renewcommand{\S}{\textsl{S}}
\renewcommand{\textsl}{\textit}
\newcommand{\llvm}{\textit{LLVM}}
\newcommand{\na}{\textsl{NA}}
\newcommand{\Cvar}[1]{\textit{#1}}
\newcommand{\Cfunc}[1]{\textit{#1()}}
\newcommand{\Croutine}[1]{\Cfunc{#1}}
\newcommand{\Cstruct}[1]{\textsl{#1}}
\newcommand{\Ctype}[1]{\textsl{#1}}
\newcommand{\Cclass}[1]{\textit{#1}}
\newcommand{\Java}{\textsl{Java}}
\newcommand{\Python}{\textsl{Python}}
\newcommand{\Julia}{\textsl{Julia}}
\newcommand{\Lisp}{\textsl{LISP}}
\newcommand{\JS}{\textsl{JavaScript}}
\newcommand{\lib}[1]{\textbf{lib#1}}
\newcommand{\dquote}[1]{``#1''}
\renewcommand{\arg}[1]{\texttt{#1}}
\newcommand{\Rop}[1]{\texttt{#1}}
\renewcommand{\citep}[1]{(\cite{#1})}
\makeatother

\begin{document}
\begin{frontmatter}

\title{Enhancing \textit{R} with Advanced Compilation Tools and Methods}
\runtitle{Enhancing \textit{R} with Advanced Compilation Tools and Methods}

\begin{aug}
\author[a]{\fnms{Duncan}~\snm{Temple Lang}\corref{}\ead[label=e1]{duncan@r-project.org}}
\runauthor{D. Temple Lang}

\affiliation{University of California at Davis}

\address[a]{Duncan Temple Lang is Associate Professor, Department of Statistics,
University of California at Davis, 4210 Math Sciences Building, Davis,
California 95616, USA \printead{e1}.}

\end{aug}

%
\begin{abstract}
I describe an approach to compiling common idioms in \textit{R} code
directly to native machine code and illustrate it with several
examples. Not only can this yield significant performance gains, but
it allows us to use new approaches to computing in \textit{R}. Importantly,
the compilation requires no changes to \textit{R} itself, but is done
entirely via \textit{R} packages. This allows others to experiment with
different compilation strategies and even to define new
domain-specific languages within \textit{R}. We use the Low-Level Virtual
Machine (\textit{LLVM}) compiler toolkit to create the native code and perform
sophisticated optimizations on the code. By adopting this widely used
software within \textit{R}, we leverage its ability to generate code for
different platforms such as CPUs and GPUs, and will continue to
benefit from its ongoing development. This approach potentially
allows us to develop high-level \textit{R} code that is also fast, that
can be
compiled to work with different data representations and sources, and
that could even be run outside of \textit{R}. The approach aims to both
provide a compiler for a limited subset of the \textit{R} language and also
to enable \textit{R} programmers to write other compilers. This is another
approach to help us write high-level \textit{descriptions} of what we
want to
compute, not \textit{how}.
\end{abstract}

%

\end{frontmatter}

\section{Background \& Motivation}
%
%

Computing with data is in a very interesting period at present and
this has significant implications for how we choose to go forward with
our computing platforms and education in statistics and related
fields. We are simultaneously (i) leveraging higher-level, interpreted
languages such as $R$, MATLAB, Python and recently Julia, (ii) dealing
with increasing volume and complexity of data,
and (iii) exploiting, and coming to terms with,
technologies for parallel computing including
shared and nonshared multi-core processors and GPUs (Graphics
Processing Units).
These challenge us to innovate and
significantly enhance our existing computing platforms and to develop
new languages and systems so that we are able to meet not just
tomorrow's needs, but those of the next decade.

Statisticians play an important role in the ``Big Data'' surge, and
therefore must pay attention to logistical and performance details of
statistical computations that we could previously ignore. We
need to think about how best to meet our own computing needs for the
near future and also how to best be able to participate in
multi-disciplinary efforts that require serious computing involving
statistical ideas and methods. Are we best served with our own
computing platform such as \R\ \citep{bib:R}? Do we
\textit{need} our own system? Can we afford the luxury of our own
system, given the limited resources our field has to develop, maintain
and innovate with that system?
Alternatively, would we be better off reimplementing
\R\ or an \R-like environment on a system that is developed and
supported by
other larger communities, for example, \Python\ or \Java? Or can we
leave it to others to build a new, fast computing environment that we
can leverage within the field of statistics?
Would ``doing it right'' give us an
opportunity to escape some of the legacy in our current systems and
position ourselves for more innovation? Would developing a new system
splinter our already small community and reduce our effectiveness in
disseminating new statistical methods as effectively as we do through
\R's excellent package mechanism?

I have wrestled with these questions for over a decade. I don't
believe there is a simple answer as to what is the best way to
proceed. It is as much a social issue as a technical one. The \R\
community is an amazing and valuable phenomenon.
There is a
large \R\ code base of packages and scripts in widespread use for
doing important science. Even if new systems do emerge and replace
\R, this will take several years. We need significant improvements
in performance to make \R\ competitive with other systems, at least for
the near future. We must improve \R's performance to allow us
to continue to deal with larger and more complex data and problems.
In this paper, I discuss one direct approach to improve the
performance of \R\ code that is extensible and enables many people to
further improve it.

The essence of the approach I am suggesting is conceptually quite
simple and emerged in numerous other languages and platforms,
around the same time we first started implementing it
for \R. The idea is that we compile \R\ code directly to
low-level native machine instructions that will run on a CPU or GPU or
any device we can
target. Instead of insisting that \R\ code be evaluated by
\textit{the} one and only \R\ interpreter, we may generate the code
to perform the equivalent computations in a quite different way. We
can dynamically compile fast native code by combining information
about the code, the data and its representation being processed by
that code, the available computing ``hardware'' (i.e., CPU or GPU or
multi-core), the location of the different sources of the data,
whether we need to handle missing values (NAs) or not, and so on.
This is a form of just-in-time (JIT) compilation. It leverages
additional knowledge about the context in which code will be run. It
maps the code to low-level machine instructions rather than having it
evaluated by a high-level interpreter.

The approach presented here is quite different from
how programmers typically improve performance for \R\ code.
They manually implement the slow, computationally intensive parts
in \C/\Cpp, and call these \mbox{routines} from \R.
I call this ``programming around \R.'' Instead, I am trying to
``\textit{compile} around \R.''
In this approach,
statisticians use familiar $R$ idioms to express the desired computations,
but the compiler infrastructure generates optimized instructions
to realize the intended computations as efficiently as possible on
the hardware platform in use.
The input from the statistician or analyst to this process is \R\ code, not
low-level \C/\Cpp\ code. This is good because humans can more
easily understand, debug, adapt and extend code written in $R$.
Furthermore, the compiler can ``understand'' what is intended
in the high-level code and optimize in quite different ways.
This also allows the code to be optimized in very different ways in
the future.
The high-level code says what to do, but not how to do it.
How is left to the compiler.


What makes this approach feasible and practical now is
the availability of the Low-Level Virtual Machine Compiler
Infrastructure (\llvm) \citep{bib:llvm}. \llvm\ is the winner of the
2012 Association for Computing Machinery System Software Award (the
same award
conferred on the \S\ language in
1999) and is a highly extensible compiler toolkit.
LLVM is a \Cpp\ library we can integrate into \R\ (and other
languages) to generate native code for various different CPUs, GPUs
and other output targets. The ability to integrate this very
adaptable and extensible tool into high-level languages is a ``game
changer'' for developing compilation tools and strategies. We can use
a technology that will continue to evolve and will be developed by
domain experts. We can adapt these to our purposes with our domain
knowledge. We do this within an extensible \R\ package rather than in
the \R\ interpreter itself. This leaves the compilation
infrastructure in ``user'' space, allowing development of any new
compilation strategies to be shared without any changes to the \R\
interpreter. This contrasts with \R's byte-code compiler and
the byte-code interpreter which is part of the fixed \R\ executable.


The \OmgPkg{Rllvm} \citep{bib:Rllvm} package provides \R\ bindings to
the \llvm\ \Cpp\ API. We can use this to generate arbitrary native code.
The \OmgPkg{RLLVMCompile} \citep{bib:RLLVMCompile} package provides a simple
``compiler'' that attempts to translate \R\ code to native code by
mapping the \R\ code to instructions in \llvm, leaving
\llvm\ to optimize the code and generate the native machine code.

Before we explore some examples of how we can use \llvm\ in \R\ to
improve performance and change our computational strategies for
certain types of problems, it is worth thinking a little about the
potential implications of fast \R\ code:
\begin{itemize}
\item\textit{Alternative data models}. On a practical level, if we
can compile scalar (i.e., nonvectorized) \R\ functions so that they
are almost as fast as \C/\Cpp\ code, we can use them to process
individual observations in a streaming or updating manner. This means
we can escape the highly-vectorized or ``all data in memory''
approaches that \R\ strongly rewards.
\item\textit{Exporting code to alternative execution environments}.
We can write \R\ code and then export it to run in
other different systems, for example,
databases, \Python, \JS\ and Web browsers.
We can map the \R\ code to \llvm's intermediate representation (IR).
We can then use \texttt{emscripten} \citep{bib:emscripten} to
compile this directly to \JS\ code. For other systems, we can share
the IR code from \R\ and they can use their own \llvm\
bindings to compile it to native code for the particular hardware.
\item\textit{Richer data structures}.
\R\ provides a small number of primitive data types, for example, vectors,
lists, functions. We can currently use these to create new aggregate
or composite data structures. However, we can only introduce new and
different data structures such as linked lists and suffix trees as
opaque data types programmed in native (\C/\Cpp) code. When we compile
code to native instructions, we also have the opportunity to have that
new code use these different data structures and to represent the data
differently from \R. The same \R\ code can be merged with
descriptions of new data types to yield quite different native
instructions that are better suited to particular problems.
\item\textit{Templating concepts}.
Our ability to create native code from \R\ code allows us to think
about \R\ functions or expressions differently. They are
descriptions of what is to be done, without the specifics of how to do
them. An \R\ compiler can rewrite them or generate code that will
behave differently from the \R\ interpreter\vadjust{\goodbreak} but give the same results
(hopefully). The functions are ``templates.'' The compiler can use
knowledge of the particular representation of the data the functions
will process to generate the native code in a more intelligent manner.
For example, the code may access elements of a two-dimensional data
set---rows and columns. There are two very different representations
of this in \R---~data frames and matrices. How the individual
elements are accessed for each is very different. The compiler can
generate specialized code for each of these and might even change the
order of the computations to improve efficiency (cache coherency) for
these representations. The function is not tied to a particular data
representation.
\end{itemize}
In summary, compiling \R\ programs through LLVM
yields novel computational potentials that are
directly relevant to improving statistical learning
and communication in the big data era.
Compiling high-level code to native code is used in many
systems. \Julia\ is an interesting modern project doing this. NumPy
\citep{bib:numpy}
in \Python\ is \mbox{another}. Several years ago, Ross Ihaka and I explored
using \Lisp\ \citep{bib:LISPBackToFuture} as the platform for a new
statistical computing
platform. The same ideas have been used there for many years and the
performance gains are very impressive.

A very important premise underlying the approach in this paper is that
the \R\ project and its large code base are important, and will be for
at least another 5 years. Users are not likely to immediately change
to a new system, even if it is technically superior. For that reason,
it is important to improve the performance of \R\ now. It is also
important to allow developers outside of the core \R\
development team to contribute to this effort and to avoid many
forked/parallel projects. For this, we need an extensible system within
\R,
and not one that requires continual changes to the centralized \R\
source code.

In addition to focusing on the immediate and near-term future and
improving \R, we also need to be exploring new language and computing
paradigms within the field of statistics. \Julia\ is an interesting
modern project doing this. We need to foster more experiments so new
ideas emerge. To do this, we also need to increase the quantity and
quality of computing within our curricula.

Section~\ref{sec:Examples} constitutes the majority of this paper. In
it I explore different examples of computing in \R\ and how
compiling code can make the computations more efficient, both\vadjust{\goodbreak}
by simply obtaining faster execution speed and also by allowing us to
change how we
approach the problem. In Section~\ref{sec:FutureStrategies} I
discuss some additional general strategies we can exploit to improve
computations in the future. I briefly discuss other exciting research
projects to improve \R's efficiency and contrast them with the \llvm\
approach. I outline in Section~\ref{sec:FutureWork} a road map of the
ongoing work on the \llvm\ approach and other related projects as
part of our future activities. The aim of this is to
illustrate the feasibility of the entire approach.

In this paper I focus on reasonably standard \R\ code and approaches
that can be improved by generating native code. The semantics of the
generated code are the same as the original \R\ code. The approach
also allows us to develop new languages and semantics within the
compilation framework and to explore different computational
models. However, the examples discussed in this paper stay within \R's
existing computational model in order to anchor the discussion and
avoid too many degrees of freedom becoming a distraction. I do hope
that we will explore new semantics and language features within \R\
via compilation.


\section{Enhancing \textit{R} with Advanced Compilation Tools and
Methods}\label{sec:Examples}

In this section we'll explore some examples of how we can write
code in \R\ and compile it to machine code. These explore different
strategies and illustrate how we can approach computations
differently when we have the option to compile code rather than only
interpret it. We
have chosen the problems for several reasons. They are each reasonably
simple to state, and they illustrate the potential benefits of
compilation. Like most benchmarks, some of the examples may not
reflect typical use cases or how we would do things in \R. However,
most of these problems are very concrete and practical, and represent
ways we
would like to be able to program in \R, were it not for the
performance issues. In this way, the examples illustrate how we can
continue to use \R\ with an additional computational model and can
overcome some of the interpreter's performance issues while still
using essentially the same \R\ code.

\begin{Note*} In the following subsections, we present absolute and
comparative timings for different approaches and implementations to
the different tasks. These timings were performed on three different
machines. We used a MacBook Pro running OS X
(2.66~GHz Intel Core i7,
8~GB 1067 MHz DDR3) and also two Linux machines.\vadjust{\goodbreak} The first of these
Linux machines is an older 2.8~GHz AMD Opteron, 16~GB. The second is a
much more recent and faster machine---3.50~GHz Intel Core i7-2700K,
with 32~GB of RAM. Additionally, the different machines have different
compilers used to compile \R\ itself and these may impact the
timings. We used GCC 4.2.1 on OS X, and GCC 4.3.4 on the first Linux
machine and both GCC 4.8.0 and clang on the second Linux box. In all
cases, \R\ was compiled with the -O3 optimization level flag. The
absolute times are quite different across machines, as we expect, and
the within-machine relative performance of our \llvm\ generated code
to native code \mbox{differs} between OS X and Linux. However, the
within-machine results are very similar across Linux machines.
Finally, our current steps to optimize the native code we generate
with \llvm\ are quite simple and we expect to improve these in the
near future.
\end{Note*}

We have not included the time to compile the code in our
measurements. There are two steps in this---compiling the $R$ code to
intermediate representation (IR) and then compiling the IR to native
code. The former can be done once, and the latter for each \R\
session and is done in \llvm's \Cpp\ code. There are several reasons
for omitting these steps in the timings. First, our focus is on
tasks that take a long time to run in \R, for example, many hours or days.
Compilation time will be on the order of, at most, a few minutes and
so the compilation time is negligible. Second, we expect that the
compiled code will be reused in multiple calls and so the overhead of
compiling will be amortized across calls. We have also ignored the
time to byte-compile $R$~functions, or compile and install \C/\Cpp\
code to be used in \R\ packages.

%
\begin{table*}[b]
\tablewidth=\textwidth
\tabcolsep=0pt
\caption{Timings for computing Fibonacci Sequence Values}\label
{tab:FibSpeedFactor}
\begin{tabular*}{\textwidth}{@{\extracolsep{\fill}}ld{2.2}d{3.2}d{3.2}d{3.1}d{2.3}d{3.1}@{}}
\hline
& \multicolumn{2}{c}{\textbf{OS X}}
& \multicolumn{2}{c}{\textbf{Linux 1}}
& \multicolumn{2}{c@{}}{\textbf{Linux 2}}
\\
\ccline{2-3,4-5,6-7}
& \multicolumn{1}{c}{\textbf{Time}}
&\multicolumn{1}{c}{\textbf{Speedup}}
& \multicolumn{1}{c}{\textbf{Time}}
&\multicolumn{1}{c}{\textbf{Speedup}}
&\multicolumn{1}{c}{\textbf{Time}}
&\multicolumn{1}{c@{}}{\textbf{Speedup}}
\\
\hline
Interpreted \R\ code & 80.49 & 1.00 & 112.70 & 1.0 & 51.780 & 1.0
\\
Byte-compiled \R\ code & 31.70 & 2.53 & 45.85 & 2.5 & 21.620 & 2.4
\\
\OmgPkg{Rllvm}-compiled code & 0.12 & 653.90 & 0.21 & 526.4 & 0.097 & 531.0
\\
\hline
\end{tabular*}
\tabnotetext[]{}{These are the timings for a call to \Rexpr{fib(30)}
using the
regular \R\ function, the byte-compiled version and the
\llvm-compiled version. To improve the accuracy of the timings, we
calculate the duration for $20$ replications for the two slower
functions and $200$ replications of the \llvm-compiled routine and
divided the duration by 10. The \llvm-compiled version is clearly
much faster.}
\end{table*}

\subsection{The Fibonacci Sequence}

The Fibonacci sequence is an interesting mathematical sequence
of integers defined by the recurrence/recursive relation
\[
F_n = F_{n-1} + F_{n-2},\quad n \ge0
\]
with $F_0 = 0$ and $F_1 = 1$.
We can implement this as an \R\ function in an easy and obvious manner
as\vspace*{7pt}
\begin{verbatim}
fib = function(n)
{
  if (n < 2L)
    n
  else
    fib(n - 1L) + fib(n - 2L)
}
\end{verbatim}
For simplicity, we don't verify that \Rvar{n} is nonnegative,
assuming the caller will provide meaningful inputs.
This maps the mathematical description of the sequence to a
computational form essentially in a one-to-one manner. This is a good
thing, as it makes the code and computations easy to understand, debug
and maintain. However, this is a scalar function and not vectorized,
meaning that it computes the value of the Fibonacci sequence for a
single integer value rather than element-wise for a vector of inputs.
This makes it slow in \R\ if we want to compute multiple values from
the sequence, for example, apply it to each element of a vector.
Instead of
implementing the function in this natural form, to gain performance,
we would look to other implementations. Since the sequence is
described by a recurrence relationship, there is a simple closed-form
formula for computing the $n$th value of the sequence which can be
easily implemented in \R\ as a vectorized function of
$n$. Alternatively, we might use memoization to remember results
computed in previous calls to the function to avoid repeating
computations. We might even use a lookup table of pre-computed
results for common input values, or some combination of these
approaches. The key is that to get good performance, we have to think
about the problem quite differently. Instead, we'll explore how we
make the simple implementation above perform better and hope to avoid
having to change our entire approach or rely on \R's other vectorized
operations.

We use the function \Rfunc{compileFunction} in the \OmgPkg{RLLVMCompile}
package to create a native compiled version of the \Rfunc{fib} function with\vspace*{7pt}
\begin{verbatim}
fib.ll = compileFunction(fib,
  Int32Type, list(n = Int32Type))
\end{verbatim}\vspace*{7pt}
We have to specify the type of the return value and also the type of
the input(s), that is, \Rvar{n} in this case. For\vadjust{\goodbreak} this function, both the
return type and the input are regular integer values corresponding to
the 32-bit integer type \Rvar{Int32Type}. We could use a 64-bit
integer by using the type \Rvar{Int64Type} if we wanted to deal with
larger numbers. In fact, we can create two separate and different
versions of this function with different types with two calls to
\Rfunc{compileFunction}. This is a simple illustration of how easy it
is to adapt the same \R\ code to different situations and create
different compiled routines with different characteristics.

\Rfunc{compileFunction} can return an \R\ function which we can
invoke directly. However, by default, it currently returns an object
representing the compiled routine in \llvm. We can invoke the routine using
this object and the \Rfunc{.llvm} function, analogous to the
\Rfunc{.Call} and \Rfunc{.C} functions in \R. So\vspace*{7pt}
\begin{verbatim}
.llvm(fib.ll, 30)
\end{verbatim}\vspace*{7pt}
calls our compiled routine and returns the value $832\mbox{,}040$.
Unlike the
\Rfunc{.Call}/\Rfunc{.C} functions, the \Rfunc{.llvm} function knows
the expected types of the routine's parameters and so coerces the
inputs to the
types expected by the routine. In this case, it converts the \R\ value
$30$ from what is a numeric value to an integer.

After verifying that the routine gives the correct results, we can
explore the performance of the code. This recursive function is very
computationally intensive. When calculating, for example,
\Rexpr{fib(30)}, we calculate \Rexpr{fib(28)} twice [once for each of
\Rexpr{fib(30 - 1)} and \Rexpr{fib(30 - 2)}] and, similarly, we
compute \Rexpr{fib(27)} multiple times and so on. This repetition is
one of the reasons the code is so slow. We'll compare the time to
evaluate \Rexpr{fib(30)} using three different versions of
the \Rfunc{fib} function: the original interpreted function, the
\llvm-compiled routine (\Rvar{fib.ll}) and a version of \Rfunc{fib} that
is compiled by \R's byte-compiler. The \llvm-compiled
routine is the fastest. Table~\ref{tab:FibSpeedFactor}\vadjust{\goodbreak} shows the
elapsed times for each and a ratio of the time for each function
relative to
the time for the interpreted function. These convey the relative
speedup factor.
We see that on a Macbook Pro, the \llvm-compiled
routine is $600$ times faster than the \R\ interpreter. On a Linux
machine, the speedup is a bit smaller, but still very
significant at a factor of $500$. While we have attempted to reduce the
variability of these timings, we have observed different speedups
ranging from $400$ to $600$ on OS X and $230$ to $540$ on Linux. The
timings we report here are the most recent (rather than the
``best'').\looseness=-1

Again, this is a simple example and not necessarily very
representative of how we would calculate the Fibonacci sequence in
production code. However, the ability to express an algorithm in its
natural mathematical form makes it easier to program, verify and
extend. We would very much like to be able to write code in this manner,
without sacrificing good run-time performance.

\subsection{2-Dimensional Random Walk}

Ross Ihaka developed a very instructive example of writing
straightforward \R\ code compared with clever, highly vectorized \R\
code as a means to illustrate profiling in $R$ and how to make code
efficient. The task is simulating a two-dimensional random walk. It
is very natural to write this as a loop with N iterations
corresponding to the N steps of the walk. For each step, we toss a
coin to determine whether we move horizontally or vertically. Given
that choice, we toss another coin to determine whether to move left or
right, or up or down. Then we calculate and store the new location.
We'll call this the na\"ive approach and the code is shown in
Figure~\ref{fig:rw2dNaive}. After several refinements based
on profiling and nontrivial knowledge of \R, Ihaka defines a very
efficient $R$ implementation of the random walk, shown in
Figure~\ref{fig:rw2dVector}. It removes the explicit
loop, samples all N steps in one call to \Rfunc{sample}, and
determines the positions using two calls to the \Rfunc{cumsum}
function. This makes very good use of several of \R's vectorized
functions which are implemented in \C\ code and therefore fast.

%
\begin{figure}
{\fontsize{9.5}{11.5}\selectfont{\begin{verbatim}
rw2d1 =
function(n = 100)
{
  xpos = ypos = numeric(n)
  truefalse = c(TRUE, FALSE)
  plusminus1 = c(1, -1)
  for(i in 2:n) {
        # Decide whether we are moving
        #  horizontally or vertically.
    if (sample(truefalse, 1)) {
        xpos[i] = xpos[i-1] +
                     sample(plusminus1, 1)
        ypos[i] = ypos[i-1]
      }
      else {
        xpos[i] = xpos[i-1]
        ypos[i] = ypos[i-1] +
                     sample(plusminus1, 1)
      }
    }
    list(x = xpos, y = ypos)
}
\end{verbatim}}}%
\caption{The na\"ive implementation of the 2-D random walk.}\label
{fig:rw2dNaive}
\end{figure}

%
\begin{figure}
{\fontsize{9.5}{11.5}\selectfont{\begin{verbatim}
rw2d5 =
# Sample from 4 directions, separately.
function(n = 100000)
{
  xsteps = c(-1,  1,  0,  0)
  ysteps = c( 0,  0, -1,  1)
  dir = sample(1:4, n - 1, replace = TRUE)
  xpos = c(0, cumsum(xsteps[dir]))
  ypos = c(0, cumsum(ysteps[dir]))
  list(x = xpos, y = ypos)
}
\end{verbatim}}}%
\caption{The fast, vectorized implementation of the 2-D random
walk.}\label{fig:rw2dVector}
\end{figure}

%
\begin{table*}
\tablewidth=\textwidth
\tabcolsep=0pt
\caption{Timings for simulating a 2-D Random Walk}\label{tab:rw2dTimings}
\begin{tabular*}{\textwidth}{@{\extracolsep{\fill}}ld{3.2}d{3.1}d{3.1}d{3.1}d{3.2}d{3.2}@{}}
\hline
& \multicolumn{2}{c}{\textbf{OS X}} & \multicolumn{2}{c}{\textbf{Linux 1}}
& \multicolumn{2}{c@{}}{\textbf{Linux 2}}
\\
\ccline{2-3,4-5,6-7}
& \multicolumn{1}{c}{\textbf{Time}} &
\multicolumn{1}{c}{\textbf{Speedup}} &
\multicolumn{1}{c}{\textbf{Time}}
& \multicolumn{1}{c}{\textbf{Speedup}}
& \multicolumn{1}{c}{\textbf{Time}} &
\multicolumn{1}{c@{}}{\textbf{Speedup}}
\\
\hline
Interpeted \R\ code & 171.08 & 1.0 & 196.6 & 1.0 & 100.3 & 1.0
\\
Byte compiled code & 123.92 & 1.4 & 120.8 & 1.6 & 60.51 & 1.66
\\
Vectorized \R\ code & 0.97 & 176.5 & 1.8 & 106.8 & 0.63 & 159.46
\\
\OmgPkg{Rllvm}-compiled code & 0.52 & 329.3 & 1.1 & 180.3 & 0.40 & 250.12
\\
\hline
\end{tabular*}
\tabnotetext[]{}{We generate $10$
million steps for each approach. We compare a manually vectorized
implementation in \R\ code with a na\"ive version written in \R,
both a byte-compiled and \llvm-compiled version of that na\"ive
function. The vectorized version is $175$ times faster than the
regular \R\ function. However, the \llvm-compiled version outperforms the
vectorized version, most likely by removing one \C-level loop.}
\end{table*}

We manually compiled the na\"ive implementation using \OmgPkg{Rllvm}
and, similarly, used $R$'s byte-compiler to create two compiled versions of
this function. We then simulated a 10 million step random walk using
each of the original na\"ive function, the byte-code compiled function, the
fully vectorized version and the \llvm-compiled version.
Table~\ref{tab:rw2dTimings} shows the relative speedups. We see that
the manually vectorized \R\ function\vadjust{\goodbreak} is 175 times faster than the
na\"ive implementation, illustrating how important vectorization is to
make \R\ code efficient. However, we also see that compiling the
na\"ive implementation with \llvm\ outperforms even the vectorized version,
taking about between 55\% to 65\% of the time of the vectorized version.
This is probably due to the compiled code using a single loop, while
the vectorized version has two
calls to \Rfunc{cumsum} and hence at least one additional \C-level
loop over the N steps.

\subsection{Sampling a Text File}\label{eg:SamplingCSV}

{\spaceskip=0.185em plus 0.05em minus 0.02em Suppose we have one or more large comma-separated} value (CSV) files.
For example, we can download airline traffic delay data for each year
as an approximately $650$ megabyte CSV file from the Research and
Innovative Technology Administration (RITA),\vadjust{\goodbreak} part of the Bureau of
Transportation, at
\url{http://www.transtats.bts.gov/DL\_SelectFields.asp?Table\_ID=236}.
Rather than working with the entire data set, we might choose to take
a random sample of the observations. (We don't concern ourselves here
with the appropriateness of a simple random sample.) We'll also
assume that we know the number of observations in the CSV file.

How do we efficiently extract a sample of the lines from the file? We
could use UNIX shell tools, but it is difficult to randomly generate
and specify the lines to sample. Sampling the indices is something we
want to do in \R, but then passing all of these to a shell command is
awkward, at best. Alternatively, we could do the entire sampling in \R.
We could read
the entire file into memory [via the \Rfunc{readLines} function] and
then subset the ones we want. However, this requires a significant
amount of memory. We first store all of the lines, then make a copy
of the ones we want and then discard the larger vector. This may not
be feasible, as we may not have enough memory, or it may simply be too
slow.

We can think of different strategies. One is to first identify the
indices of all of the lines we want in our sample, and then read the
file in chunks until we get to a line that is in our sample. We store
that line and continue to read from where we are up to the next line
in our sample, and so on. To make this work, we need to be able to
continue to read from where we currently are in the file. We can use
an \R\ file connection to do this.

Our first step is to generate the vector of the line
numbers we want to sample using, for example,\vspace*{7pt}
\begin{verbatim}
lineNum = sort(sample(1:N,
  sampleSize))
\end{verbatim}\vspace*{7pt}
where \Rvar{N} is the number of lines in the CSV file.
We have sorted the line numbers,\vadjust{\goodbreak} as we will read the sample lines in
the file sequentially.
The next step is to determine how many lines there are between
successive lines in our sample. We can compute this in \R\ with\vspace*{6pt}
\begin{verbatim}
lineSkips = diff(c(0, lineNum))
\end{verbatim}\vspace*{6pt}
which gives a vector of the pairwise difference between successive elements.
For example, suppose the first two lines we want to sample are 60 and
200. The first two elements in \Rvar{lineSkips} will be $60$ and
$140$. We can then read the first two lines in our sample with\vspace*{6pt}
\begin{verbatim}
con = file("2012.csv", "r")
readLines(con, 60)[60]
readLines(con, 140)[140]
\end{verbatim}\vspace*{6pt}

Each element of \Rvar{lineSkips} tells us how many lines to read to get
the next line in our sample. So next we need a function that can read
that many lines and return the last of these.
The following function does\vspace*{6pt} this:
\begin{verbatim}
readTo = function(numLines, con)
  readLines(con, numLines)[numLines]
\end{verbatim}\vspace*{6pt}

The final step to obtaining our entire sample is to call
\Rfunc{readTo} for each element of \Rvar{lineSkips}, for example,\vspace*{6pt}
\begin{verbatim}
readSelectedLines =
  function(lineSkip, file)
  sapply(lineSkip, readTo, file)
\end{verbatim}\vspace*{7pt}
To obtain our sample, we call \Rfunc{readSelectedLines}, passing it the
variable \Rvar{lineSkips} and our open connection:\vspace*{7pt}
\begin{verbatim}
con = file("2012.csv", "r")
sample =
  readSelectedLines(lineSkips, con)
\end{verbatim}

Each of these functions is concise and efficient since \Rfunc{sapply}
is essentially implemented as a \C-level loop within the \R\
interpreter. Using the connection and \Rfunc{readLines} to read
blocks of lines in \Rfunc{readTo} is efficient as it uses \C\ code
within \R. Unfortunately, it does involve reading, allocating,
storing, subsetting and discarding a potentially large character
vector returned by each call to \Rfunc{readLines}. However, we only
want a single line at the end of that vector in each call. While each
call involves significantly fewer lines than reading the entire file,
allocating a large character vector still slows the computations, as it
extensively involves the memory manager in \R. A~different approach
to avoid the memory issue is to change the \Rfunc{readTo} function so
that it reads
each line individually and then returns the last one. We could change
it to\vspace*{7pt}
\begin{verbatim}
readTo =
function(numLines, con)
{
  ans = ""
  for(i in 1:numLines)
    ans = readLines(con, 1)
  ans
}
\end{verbatim}\vspace*{7pt}
Again, this is straightforward and easy to understand. Unfortunately,
it is extremely slow as we are now looping in \R\ over almost every line
in the file.

The idea of reading one line at a time would work well if we could
avoid the overhead of the \R\ loop mechanism. We can do this if we
compile this new version of \Rfunc{readTo} into native code. We can
almost do this now, but we need to have an equivalent of \Rfunc{readLines}
to read a single line of a file. This is exactly what the standard
\C\ routine \Cfunc{fgets} does. Similar to a connection, we pass \Cfunc{fgets}
a pointer to an opaque \C-level \Cstruct{FILE} data structure, and it
puts the contents of the next line it reads into a location in memory
that we
also provide. For simplicity of exposition, we will define our own function
\Rfunc{Fgets} in \R\ as a proxy to call \Cfunc{fgets} with\vspace*{7pt}
\begin{verbatim}
Fgets = function(file)
  fgets(ptr, 1000L, file)
\end{verbatim}\vspace*{7pt}
This is \R\ code and it just assumes there is a function
named \Rfunc{fgets} and that \Rvar{ptr} is somehow
(the address of) an array in memory with $1000$ character elements,
that is, space for a long string.
We won't run this code in \R, so these variables [\Rvar{ptr},
\Rvar{file} and \Rfunc{fgets}] don't actually have to
exist in \R. Instead, we will allocate them in \llvm\ for the
compiled, native routine we generate from \Rfunc{Fgets}.

We compile the \Rfunc{Fgets} function in an \llvm\ module, a
collection of routines and variables, using \Rfunc{compileFunction}.
We also define the module-level \dquote{global} variable \Cvar{ptr} to
be the pointer to the array we want, after creating the actual array
of $1000$ characters as another global variable. When compiling
\Rfunc{Fgets}, we also need to tell the compiler about the signature
of the external \Cfunc{fgets} routine so that it can make the call to
\Cfunc{fgets} correctly. We do this via\vspace*{7pt}
\begin{verbatim}
mod = Module()
FILEType = pointerType(Int32Type)
declareFunction(list(StringType,
  StringType, Int32Type, FILEType),
  "fgets", mod)
\end{verbatim}\vspace*{7pt}
[While we have done this explicitly, we could automate this step using
the \OmgPkg{RCIndex} \citep{bib:RCIndex} package to obtain the
signature programmatically.]
We also need to tell the \llvm\ run-time engine how to locate
the \Cfunc{fgets} routine which we do with\vspace*{7pt}
\begin{verbatim}
llvmAddSymbol("fgets")
\end{verbatim}\vspace*{7pt}

Note that in our \Rfunc{Fgets} function, we assumed that the longest
line was less than $1000$ characters. We can specify a different
length if we knew or suspected otherwise. Similarly, we didn't provide
any error checking about whether we had reached the end of the file.
This is because we are assuming that the caller knows the total number
of lines and is sampling only up to, at most, that number. This is an
example of the context-specific shortcuts we can make when compiling
the code for a particular situation and not writing general, robust
code which can be used in many different situations. We could also
tell the compiler to add these tests for us, if we wanted, but can
avoid the extra computations when we know they are redundant.

%
\begin{table*}
\tablewidth=\textwidth
\tabcolsep=0pt
\caption{Timings for sampling a CSV file}\label{tb:SampleCSVTimings}
\begin{tabular*}{\textwidth}{@{\extracolsep{4in minus 4in}}ld{2.3}d{2.2}d{3.2}d{2.1}d{2.2}d{2.1}@{}}
\hline
& \multicolumn{2}{c}{\textbf{OS X}} & \multicolumn{2}{c}{\textbf{Linux
1}}
& \multicolumn{2}{c@{}}{\textbf{Linux   2}}
\\
\ccline{2-3,4-5,6-7}
& \multicolumn{1}{c}{\textbf{Time}} &
\multicolumn{1}{c}{\textbf{Speedup}} &
\multicolumn{1}{c}{\textbf{Time}} & \multicolumn{1}{c}{\textbf{Speedup}}
& \multicolumn{1}{c}{\textbf{Time}} & \multicolumn{1}{c@{}}{\textbf{Speedup}}
\\
\hline
Interpreted \R\ loop \& \Rfunc{readLines} & 68.93 & 1.0 &
103.25 & 1.0 & 42.78 & 1.0
\\
\OmgPkg{Rllvm}-compiled loop \& \Cfunc{Fgets} &
3.278 & 21.09 & 6.54 & 15.8 & 2.59 & 16.5
\\
\C\ code (\OmgPkg{FastCSVSample}) & 3.0 & 22.97 & 6.28
& 16.4 & 2.40 & 17.8
\\
\hline
\end{tabular*}
\tabnotetext[]{}{We use vectorized code in \R\ to read blocks of data
and extract the
final line of each block. The \llvm\ approach compiles simple \R\
functions that read one line at a time. The \OmgPkg{FastCSVSample}
does the same thing with manually written \C\ code.
The compiled approaches avoid the memory usage related to
\Rfunc{readLines} and see a nontrivial speedup.
The \C-code in the \OmgPkg{FastCSVSample} package outperforms the
\llvm-compiled version, but both approaches outperform the approach
using \R's connections and \Rfunc{readLines} functionality
which are also implemented with \C\ code.}
\end{table*}

How do we obtain the instance of the \Ctype{FILE} data type to pass to
the compiled \Rfunc{Fgets} routine? We can use the \C\ routine
\Cfunc{fopen} and again, we can write an $R$ function that mimics that
and then compile it. However, the \OmgPkg{RLLVMCompiler} package has a
function to automate the creation of that proxy function in \R, if
we know the signature of the \C\ routine of interest. So this
example illustrates how we can dynamically create bindings to existing
compiled routines in different libraries. In the case of
\Ctype{FILE}, we can also use the existing function \Rfunc{CFILE} in
the \OmgPkg{RCurl} \citep{bib:RCurl} package.

So now we can read a single line from an open \Ctype{FILE} object in
\R\ via our compiled \Cfunc{Fgets} routine. We can redefine our
\Rfunc{readTo} function as\vspace*{7pt}
\begin{verbatim}
readTo =
function(numLines, con)
{
  ans = ""
  for(i in 1:numLines)
    ans = Fgets(con)
  ans
}
\end{verbatim}\vspace*{7pt}
This is almost identical to the original function above but
replaces the call \Rexpr{readLines(con, 1)} with \Rexpr{Fgets(con)}.
Now we can compile this into native code via
\Rfunc{compileFunction} and the resulting code will be quite fast.

We now have a fast replacement for reading up to the next line in our
sample. The last step is to make \Rfunc{readSelectedLines} fast.
Recall that this was implemented simply as \Rexpr{sapply(lineSkip,
readTo, file)}.
When we compile this as returning an \R\ character
vector, our compiler recognizes the \Rfunc{sapply} call and converts
this into a loop in native code and populates and returns a new
\R\ character vector.

In summary, we have compiled three \R\ functions [\Rfunc{Fgets},
\Rfunc{readTo} and \Rfunc{readSelectedLines}] and these now allow us
to read one line at a time and use the minimal amount of memory to
collect the lines for our sample, but using two loops in native code
rather than in \R.

We can now compare the performance of our \R-based approach using
\Rfunc{readLines} to consume chunks of lines and our compiled version
that reads one line at a time. In addition to these two approaches,
we also have a manual \C\ implementation essentially equivalent to
our \llvm-compiled approach in the \OmgPkg{FastCSVSample} package
\citep{bib:FastCSVSample}. Our
timings are based on extracting a sample of one hundred thousand lines
uniformly from a CSV file that contains one hundred million lines---the
same lines for each approach. The elapsed times are given in
Table~\ref{tb:SampleCSVTimings}.
We see that our compiled approach of reading one line at a time is
around twenty times faster than collecting many unnecessary lines with
\Rfunc{readLines} and looping in \R, even with \Rfunc{sapply}. The
difference between the \llvm\ and native \C\ approaches may be
inherent, but also possibly due to different optimization techniques
that we may be able to enhance with \llvm. In short, we can
outperform \R's native vectorized code by compiling our
relatively straightforward \R\ code.

The exposition of this example may make it seem more complicated than
it is. Essentially, we want to efficiently read one line of a file at
a time in order to get to the next line in our sample. We compiled
the \Rfunc{Fgets} function for this and then compiled two other
functions in \R\ to perform loops over the number of lines. The
important implications from this example is that we can sidestep $R$'s
memory management, get fine-grained control over computations using
dynamically generated routines, and we can use existing native
routines and data structures, such as \Cfunc{fgets} and \Ctype{FILE},
in our \R\ code that will be compiled. We could already dynamically
call native routines directly from \R\ using, for example,
\Rpkg{rdyncall} \citep{bib:rdyncall} or \Rpkg{Rffi} \citep{bib:Rffi}.
What is important here is that we are also compiling the iterations
and not doing them in $R$.

\subsection{Fusing Loops}

Consider computing the log-likelihood for a given vector of
observations \Rvar{x} and a density function, say, \Rfunc{dnorm}.
In \R, we can write the log-likelihood very efficiently as\vspace*{7pt}
\begin{verbatim}
sum(log(dnorm(x, mu, sigma)))
\end{verbatim}\vspace*{7pt}
Indeed, we could reduce this to \Rexpr{sum(dnorm(x, mu, sigma, log = TRUE))},
but the purpose of this example is to consider a general sequence of
calls to vectorized \R\ functions.

Each of the functions \Rfunc{dnorm}, \Rfunc{log} and \Rfunc{sum} are
built-in to \R\ and are implemented in \C\ code, and two of them use
the very efficient \Rfunc{.Primitive} mechanism. As a result, this
code seems to be as fast as it can be. This is true given the way
\R\ interprets the expression, one sub-expression at a time.
However, there are two ways we can
make this more efficient by compiling such an expression. The first
is by reducing the number of loops (in \C) from three to one.
Generally, if we have $n$ nested-calls to element-wise and vectorized
functions, we can reduce $n$ loops to one. Second, we can typically
eliminate at least one allocation of a potentially large vector. A
third way we might speed up the computations is to use parallel
capabilities such as a GPU or multiple cores or CPUs. We won't discuss
this here, but it is conceptually quite straightforward to do generally
when we are
compiling the code dynamically. Indeed, the ability to
programmatically combine a particular function with a general
parallel strategy makes it more expedient than writing it ourselves in
\C/\Cpp.

How does \R\ evaluate the expression above? It uses three separate
loops. Ignoring pedantic details, essentially \R\ evaluates the call
to \Rfunc{dnorm} and so loops over all of the elements of \Rvar{x} and
computes the density at each of those values. It stores these values
in a newly allocated vector and then returns that. This becomes the input
to the call to \Rfunc{log}. \R\ then iterates over the elements of
this vector and computes the \Rfunc{log} for each individual value.
In this case, \R\ may recognize that it doesn't need to create a new
vector in which to return the results, but that it can reuse the input
vector since it is essentially anonymous.
The final step in the overall expression is the call to \Rfunc{sum}
and this iterates over the elements of the vector it receives and
returns a single
scalar value.

%
%

Importantly, there are three loops over three vectors all of the same
length, and we allocate one new and large vector.\vadjust{\goodbreak}
We could use a single loop and avoid allocating this intermediate
vector by rewriting the code as\vspace*{7pt}
\begin{verbatim}
normalLogLik =
function(x, mu = 0, sigma = 1) {
  ans = 0
  for(val in x)
    ans = ans +
      log(dnorm(val, mu, sigma))
  ans
}
\end{verbatim}\vspace*{7pt}
Instead of the vectorized calls in \R, we have put scalar function
calls inside a single loop. We have combined the calls to
\Rfunc{dnorm} and \Rfunc{log} together. Then we took the result for
each element and added it to the cumulative sum. This combination of
operations is
called loop fusion and for large vectors can yield significant
performance improvements.

%
\begin{table*}
\tablewidth=13cm
\tabcolsep=0pt
\caption{Times and relative performance for fusing loops}\label{tab:fuseTimings}
\begin{tabular*}{13cm}{@{\extracolsep{\fill}}lccc@{}}
\hline
& \textbf{OS X} & \textbf{Linux 1} & \textbf{Linux 2}
\\
\hline
\OmgPkg{Rllvm}-compiled fused loops & 0.73 & 1.69 & 0.84
\\
Interpreted \R\ vectorized functions & 0.52 & 2.28 & 1.07
\\[3pt]
Regular \R\ time/\OmgPkg{Rllvm} time & 0.71 & 1.35 & 1.27
\\
\hline
\end{tabular*}
\tabnotetext[]{}{The first two rows show the times for fusing the loops
by compiling the \R\ with \llvm\ and using a sequence of calls to
\R's vectorized functions.
The final row shows the ratio of the two times within each machine.
Fusing the loops is slower on OS X, but faster on Linux.}
\end{table*}

This new scalar version is faster by avoiding the loop and allocation.
Of course, it is evaluated in \R\ and so will be much slower. We
could write this in \C, but it would be very specific to the
log-likelihood for a \mbox{Normal} density. Generally, we would
have to
write implementations for various sequences of calls, for example, for
different density functions [i.e., \Rexpr{sum(log(pdf(x, ...)))}], and
expressions involving other functions [e.g., \verb+prod(dchisq(x^2,+ \verb+p))+].
This isn't practical. However, given our ability to dynamically\vadjust{\goodbreak}
generate native code, we can compile any expression such as our
original expression
\Rexpr{sum(log(dnorm(x, mu, sigma)))} into the native equivalent of our
scalar code above.

To compile the \Rfunc{normalLogLik} function above, we need to be able
to call scalar versions of the \Rfunc{log} and \Rfunc{dnorm} routines.
The \Rfunc{log} function is available in the ubiquitous math library
(\lib{m}) and we can just refer to it. The Normal density function is
not standard. We can arrange for our native code to invoke \R's
\Rfunc{dnorm} function for each scalar value in the vector. This is
both awkward and inefficient. Instead, we can write our own version
of \Rfunc{dnorm} directly in \R. While this would be slow to invoke
many times in \R, we will compile our \Rfunc{dnorm} and
\Rfunc{normalLogLik} functions together into a single module and both
will be fast. Another possible approach, in this case, is to take
advantage of the good design and modularity of the Rmath library. It
provides the routine \Croutine{dnorm4} as a regular native
routine (unconnected with \R's data types, etc.) and so we can invoke
it, just as we do the \Croutine{log} routine.

For reasons that are not quite clear at present, on the OS X machine,
our loop-fused version\vadjust{\goodbreak} takes about $40\%$ longer than the
\R\ code for 10 million observations and $27\%$ longer for 100
million observations. Again, we suspect that we will be able to
improve the \llvm-compiled code by exploring more of its optimization
facilities. However, on the Linux machines, we do see a speedup, even
for 10 million observations where the \llvm\ loop-fused code runs in
about $75\%$ the time of the \R\ code.
The timings and relative performances are given in
Table~\ref{tab:fuseTimings}.
Regardless of the exact
numbers, the results indicate that compiling our own code is
competitive with manually writing vectorized routines in \R,
and that we can outperform these built-in \C\ routines.

A difference between the two approaches is that \R\ uses the
\Rfunc{.Primitive} mechanism rather than a standard function call
which we have to do via the \Rfunc{.llvm} function. However, not only
do we reduce three loops to one, but we also avoid dealing with
missing values (\na{}s) and additional parameters such as \arg{base}
for the
\Rfunc{log} function. So we should be doing even better. If we have
access to multiple cores or GPUs, we may be able to execute this code
much more efficiently simply via parallel execution.
By fusing the loops operations together, we can also avoid three
separate transitions from the host to the GPU and transferring memory
between the two systems more times than we need.

We explicitly wrote the \Rfunc{normalLogLik} function
to show how to\vadjust{\goodbreak} fuse the loops.
We could also have written the original expression
\Rexpr{sum(log(dnorm(x, mu, sigma)))} as\vspace*{7pt}
\begin{verbatim}
Reduce(`+`, Map(log, Map(dnorm, x,
  MoreArgs = list(mu, sigma))))
\end{verbatim}\vspace*{7pt}
By explicitly using these functional programming \mbox{concepts}, it is easy
for us
to see how to fuse loops and rewrite the code into the loop above.
The \OmgPkg{RLLVMCompile} package can recognize such an expression
and compile it to the loop-fused instructions.
We can either require \R\ programmers to do this in order to gain the
performance from native code or we can try to make the compiler
recognize the vectorized nested function call idiom of the form
\Rexpr{f(g(h(x)))}.


\subsection{Computing Distances between Observations}

Distances between pairs of observations are important in common
statistical techniques such as clustering, multi-dimensional scaling,
support vector machines and many methods that use the ``kernel
trick'' \citep{LearningWithKernels}. \R\ provides the \Rfunc{dist}
function that allows us to
compute the distance between all pairs of observations in a matrix or
data frame, using any of six different metrics. The core computations
are implemented in \C\ and are fast. However, there are some issues
and rigidities.

The \Rfunc{dist} function insists that the data passed to the \C\
code are represented as a matrix, and so will make a copy of the data
if a data frame is given by the caller. For large data sets, this can
be a significant issue as we will essentially have two copies of the
data in memory. Also, the \Rfunc{dist} function only accepts a single
data set and computes the distances between all pairs of observations
within it. In contrast, a reasonably common situation is that we start
with two separate data sets---\Rvar{X} and \Rvar{Y}---and want to
compute the distance between each observation in \Rvar{X} and each
observation in \Rvar{Y}, but not the distances between pairs of observations
within \Rvar{X} or within \Rvar{Y}.
Not only do we risk having three copies of
the data in memory (the two separate data frames, the two combined
into one data frame and then converted to a matrix), but the \Rfunc
{dist} function
will also perform many unnecessary computations for\vadjust{\goodbreak} these within-same-set
observations that we will discard. If we have two data sets
with $n1$ and $n2$ observations, respectively, the \Rfunc{dist}
function computes $(n1 + n2)\times(n1 + n2 - 1)/2$ distances. We are
only interested in $n1\times n2$ of these. As $n1$ and $n2$ diverge,
the number of unnecessary computations increases, and this is especially
burdensome if the number of variables for each observation is large.

Another rigidity is that the choice of distance metric is fixed.
If we wanted to introduce a new distance metric, it would be useful to
be able to reuse the \C\ code underlying \Rfunc{dist}. We could do
this with a function pointer in \C, but the code for \Rfunc{dist}
would need to be modified to support this. Accordingly, if we want to
introduce a new metric, we have to copy or re-implement the entire \C\
code.

The \C\ code underlying \Rfunc{dist} can use parallel capabilities
(OpenMP) if they are detected when \R\ is compiled. We cannot use
GPUs or change the parallel strategy within an \R\ session without
rewriting the \C\ code. As a result, we would like to be able to
express the computations in \R\ and select a different strategy for
parallelizing the computations at run-time.

In short, as useful as \Rfunc{dist} is, we would like it to be much
more flexible. We want to be able to compute the distances between two
sets of observations, not within a single data set; use a data frame
or a matrix or perhaps some other data representation without making a
copy of the data; introduce new metrics within the same
infrastructure; and use different parallel computing approaches.
The current \Rfunc{dist} function in \R\ cannot help us meet these
goals and is
essentially static/fixed code.

The package \Rpkg{pdist} \citep{bib:pdist} provides a way to compute
pairwise distances
between two data sets. This avoids the redundant
computations. Unfortunately, it only supports the Euclidean metric and
also insists on matrices being passed to the \C\ code. Also, it has
no support for parallel computing.

If we could write the basics of the \Rfunc{dist} function in \R\ and\vadjust{\goodbreak}
make it fast, we could address all of the enhancements we listed above
as well as make the code more comprehensible and accessible to users.
The basic approach to computing the distance between each pair of
observations in two data sets \Rvar{X} and \Rvar{Y} can be expressed
in \R\ with the following quite specific/rigid function (written to aid
compiling):\vspace*{7pt}
\begin{verbatim}
dist =
function(X, Y, nx = nrow(X),
  ny = nrow(Y), p = ncol(X))
{
  ans = numeric(nx * ny)
  ctr = 1L
  for(i in 1:nx) {
    for(j in 1:ny) {
      total = 0.0
      posX = i
      posY = j
      for(k in 1:p) {
        total = total +
          (X[posX] - Y[posY])^2
        posX = posX + nx
        posY = posY + ny
      }
      ans[ctr] = sqrt(total)
      ctr = ctr + 1L
    }
  }
  ans
}
\end{verbatim}\vspace*{7pt}
The basic steps are to loop over each observation in the first
data set (\Rvar{X}) and then to loop over each observation in the
other data set (\Rvar{Y}). For each pair of observations, we compute
the distance between them via the third nested loop. We could have
made this simpler (and more general) by using a vectorized \R\
expression or calling a function to do this final loop. However, we have
\textsl{inlined} the computations directly for a reason. Suppose we
had written this part of the computation as \verb+(X[i,] - Y[j,])^2+.
Unfortunately, in \R, this would cause us to create two new
intermediate vectors, one for each of the specific rows in the two data
sets. This is
because the row of each data set is not a simple vector containing the
elements of interest which we can pass to the subtraction
function (\Rop{-}). Instead, we have to arrange the data in each row of the
matrix or data frame into a new vector of contiguous values. This is
where \R\ is convenient, but inefficient. This does not happen in
the \C\ code for \R's builtin \Rfunc{dist} routine, or ours, as it uses
matrices and knows how to
access the elements individually rather than creating a new temporary
vector. We use this same approach in our loop.
We also could allocate the vectors for the row values just once and
reuse them for each
observation, but we still have to populate them for each different observation.

%
\begin{table*}
\tablewidth=15cm
\tabcolsep=0pt
\caption{Timings for computing pair-wise distances}\label{tab:DistTimings}
\begin{tabular*}{15cm}{@{\extracolsep{\fill}}ld{2.2}d{2.2}d{2.2}@{}}
\hline
& \multicolumn{1}{c}{\textbf{OS X}} &
\multicolumn{1}{c}{\textbf{Linux   1}} &
\multicolumn{1}{c@{}}{\textbf{Linux   2}}
\\
\hline
\OmgPkg{Rllvm}-compiled code & 8.72 & 11.94 & 6.22
\\
\R\ \Rfunc{dist} function (calling native code) & 14.74 & 79.65 & 27.37
\\[3pt]
Speedup factor & 1.69 & 6.67 & 4.4
\\
\hline
\end{tabular*}
\tabnotetext[]{}{This shows the total elapsed time for distance
computations with $40$
variables and $8000$ and $1000$ observations in the two data sets.
In the \R\ approach with the \Rfunc{dist} function, there is extra
memory allocation and also $80$\% of the distances computed are discarded.
We outperform the native \R\ implementation on both platforms.}
\end{table*}

To avoid the intermediate vectors, our code explicitly accesses the
individual elements \Rexpr{X[i, k]} and \Rexpr{Y[j, k]} directly. A
matrix in \R\ is merely a
vector with the elements of the matrix arranged sequentially in column
order. Therefore, the first element of observation $i$ in \Rvar{X} is
at position $i$ in the vector. The second element of the $i$th
observation is at position \Rexpr{i + nrow(X)}, and so on. To compute the
distance between the two observations, we loop over the $p$ variables
present in each of the observations and compute the difference.

The code illustrates these computations for the Euclidean distance. We
could easily change this to implement other distance metrics. We
could do this by changing the code either manually or programmatically
by replacing the expression \verb+(X[posX]-+ \verb+Y[posY])^2+ with, for
example, \Rexpr{abs(X[posX] - Y[posY])}. Rewriting code
programmatically is a powerful feature that allows us treat \R\ code
as a template.


We can compile this three-level nested loop \R\ code via
\OmgPkg{RLLVMCompile} to native instructions. Our compiler
currently works primarily with primitive data types
and has limited support for working directly with \R\ objects,
for example, knowing the dimensions of an \R\ matrix.
Accordingly, we arrange to pass the
matrices and their dimensions to the routine
and currently have to explicitly specify the signature:\vspace*{7pt}
\begin{verbatim}
distc = compileFunction(dist,
  REALSXPType,
  list(X = DoublePtrType,
    Y = DoublePtrType,
    nx = Int32Type, ny = Int32Type,
    p = Int32Type))
\end{verbatim}\vspace*{7pt}
In the future, we will allow the caller to specify just the two data
sets (\Rvar{X} and \Rvar{Y}). However, we are making the
representation as matrices more explicit here, which is valuable
information for the compiler.

Now that we have the native code, we can then compare this to using
\R\ code that computes the same distances but does so by combining the two
data sets, calls \Rfunc{dist}, and converts the result to the
sub-matrix of interest. This comparison favors our code since this is
the form of our inputs and the expected form of the output. However,
these are quite reasonable. We timed the functions to compute the
distances for two data sets of size $8000$ and $1000$ observations,
each with $40$ variables. In this case, 80\% of the distances computed
using \R's
\Rfunc{dist} function are irrelevant and discarded. Table~\ref{tab:DistTimings}
shows the results and illustrates that by doing fewer computations, we
do indeed outperform the native \C\ code in \R, on both platforms.
If we had used data sets with similar numbers of observations,
the results would have been less dramatic. However, with $3000$
observations in \Rvar{Y}, the \llvm-generated native code was
still three times faster on Linux and only 18\% slower on OS X.

Comparing the results above to similar native code in the \Rpkg{pdist}
package, the timings again show that native \C\ code in \Rpkg{pdist}
outperforms our \llvm-compiled code, $60\%$ faster on one machine and
$9$ times faster on another. This illustrates that there is room for
significant improvement in our \llvm\ compilation. However, the fact
that we can outperform \R's native approach is encouraging. That we
can readily adapt this to different purposes and different
computational strategies indicates significant opportunities and
potential.

As a final note, we could remove the third loop and insert a call to a
function to compute the distance for these two variables,
for example,
{\spaceskip=0.2em plus 0.05em minus 0.02em \Rexpr{euclidean(X[i,], Y[j,])}. The compiler could} recognize that
\Rvar{X} and \Rvar{Y} are matrices and arrange for the compiled
version of the \Rfunc{euclidean} function to access the elements as we
have displayed above, that is, without computing the intermediate vector
for each row. If we tell the compiler \Rvar{X} and/or \Rvar{Y} are
data frames, it would generate different code to access the elements
so as to avoid these intermediate vectors. Since the compiler has the
opportunity to compile both the code for the main loop and for the
metric function together and knows the representations of the inputs,
it can create better code than if we wrote these separately and more
rigidly.



\section{Possible Compilation Enhancement Strategies}\label
{sec:FutureStrategies}

The examples in the previous section explored different ways we could
change the way we compute in \R\ with new facilities for generating
native code. We considered compiling \R\ code to native routines,
reusing existing native routines within these generated routines, and
changing the computational strategies we employ within \R\ to embrace
these new approaches. There are many other simple examples we could
consider to improve the performance of \R\ code. One is the ability
to write functions that focus on scalar operations and then to create
vectorized versions of these automatically. Given a scalar function
\Rfunc{f},
we can write a vectorized version as \Rexpr{sapply(x, f, ...)} or
with \Rfunc{mapply}. The compiler can then turn this into a native
loop. Indeed, many of the performance gains are achieved by making
looping faster. They also potentially reduce the necessity to use
vectorized code in \R\ and so hopefully make programming in \R\
more intuitive for new users.

In addition to handling loops, there are several other aspects of \R's
evaluation model that we might be able to improve by choosing different
compilation strategies in different contexts. The idea is that the
\R\ user compiling the code may have more information about the
computations, the data and its representation, or the available
computing resources than
the compiler does by examining the code. This extra information is
important. The programmer may be able to give hints to a compiler, or
choose a different compiler function/implementation altogether, to
control how the code is understood and the native instructions are
generated. The following are some reasonably obvious and general
improvements we might be able to infer or make in certain situations.
Guided by the \R\ user, different compilers may yield different
code, and even different semantics, for the same code.

\subsection{Omitting Checks for \textit{NA} Values}

Many of the \C\ routines in \R\ loop over the elements of
a vector and must check each element to see if it is a missing value
(\na).
This code is general purpose code and so this test is a fixed part
of almost every computation involving that routine.
However, when we dynamically generate the code, we may know that there
are no missing values in the data set on which we will run that code and
so omit the code to perform these additional, redundant tests.

Similarly, in our example of sampling a CSV file, we knew the number
of lines in a file and we knew that each call to the \Cfunc{fgets}
routine would succeed. As a result, we did not have to check the
return status of the call for reading at the end of the file. We also
assumed that the largest line was less than $1000$ characters and
didn't validate this in each iteration. The same applies when we are
accessing elements of a vector as to whether we first need to check
that the index is within the extent of the array or not, that is, bounds
checking. When we can verify this conceptually (within a loop over a
vector), or by declaration by the user, we can omit these checks.

These tests are typically simple and not computationally expensive.
However, they can become significant when the instructions are invoked very
often, for example, in a loop over elements of a large vector.

\subsection{Memory Allocation}

In our example discussing loop fusion, we saw that not only could we
reduce the number of overall iterations in a computation, but
we also could reduce
memory usage. We avoided creating a vector for the result of the call
to \Rfunc{dnorm} [and \Rfunc{log}]. There are potential opportunities
to further reduce memory usage.

\R\ uses the concept of pass-by-value in calls to functions. In
theory, \R\ makes a copy of each argument in a call to a function.
(Lazy evaluation means that some arguments are never evaluated and so
not copied.) However, the \R\ interpreter is smarter than this and
only copies the object when it is modified, and only if it is not part
of another object.
When compiling \R\ code, we want to be able to determine that an object
is not modified and avoid
copying it. By analyzing code, we can detect whether parameters can
be considered read-only and so reduce memory consumption in cases
where \R\ cannot verify that it is safe to avoid copying an object.
We can identify this within regular \R\ code, however, we
would have to modify the interpreter to make use of this information.
When generating native code with, for example, \Rfunc{compileFunction},
we can make use of this information dynamically,
bypassing the \R\ interpreter.

Another example where we can reduce the memory footprint of
code is when we can reuse the same memory from a different computation.
For example, consider a simple bootstrap computation something like the
following \R\ pseudo-code:\vspace*{7pt}
\begin{verbatim}
for(i in 1:B) {
  d.star = data[sample(1:n, n,
    replace = TRUE), ]
  ans[[i]] = T(d.star, ...)
}
\end{verbatim}\vspace*{7pt}
In \R's computational model, we will allocate a new data frame \Rvar
{d.star} for
each bootstrap sample. This is unnecessary. We can
reuse the same memory for each sample, as each sample has the same
structure and only differs in the values in each cell. By analyzing
the sequence of commands rather than executing each one separately
without knowledege of the others, we can take advantage of this
opportunity to reuse the memory. We can also reuse the same vector to
store the result of the repeated calls to \Rfunc{sample}. It is
reasonably clear that this is what we would do if we wrote this code
in \C, reusing the same data structure instances. However, this is not
possible within \R, as the individual computations are not as connected
as they are in the large-picture \C\ code. When we dynamically
generate native code, we can utilize this large-scale information.

Similarly, some \R\ scripts create a large object, perform several
computations on it and then move to other tasks. Code analysis can
allow us to identify that the object is no longer being used and so we
can insert calls to remove the object. However, we may be able to
recognize that the object is no longer needed, but that subsequent
tasks can reuse the same data format and representation. In that case,
we can reuse the memory or at least parts of it.

\subsection{Data Representations}

The small number of fundamental data types in \R~makes computational
reasoning quite simple,  both to use and to implement. Of course, the
choice of data type and structure can be important for  many
computations. Sequences, for example, \Rexpr{1:n} or  \Rexpr{seq(along~=
x)}, are common in \R\ code
and these are represented in \R\ as explicit vectors containing all
of the values in the sequence. We have seen that we can avoid
creating the sequence vector and populating it when it is used as a loop
counter. Similarly, we can represent a regular sequence with the
start, end and stride, that is, the increment between elements. When
generating the code, we then access elements of such a sequence using
appropriate calculations specialized to that sequence type.

In many cases, \R's simple data types cause us to use an integer when
we only need a byte, or even just a few bits, to represent a few
possible values/states. The \BiocPkg{snpStats} \citep{bib:snpStats}
package does this successfully using bytes to reduce the memory
footprint for large genomic data. Again, the operations to subset
data in this different format need to be modified from the
default. Doing this element-wise in \R\ is excessively slow. However,
when we generate native code, we are free to use different ways to
access the individual elements. This idea is important. We specify
what to do in the code, but not precisely how to do it. When
generating the code, we combine the code and information about how to
represent the data and generate different code strategies and
realizations. This is somewhat similar to template functions in \Cpp,
but more dynamic due to run-time compilation/generation with more
contextual information.

There are several other aspects of \R\ code that we can compile,
for example, matching named arguments at compile time rather than at
run time.

%
%

\section{Overview of Generating Code with \textit{LLVM}}\label
{sec:CodeGenOverview}

In this section we will briefly describe the basic ideas of how we
generate code with \llvm, \OmgPkg{Rllvm} and
\mbox{\OmgPkg{RLLVMCompile}}. This is a little more technical and low level
than our examples and readers do not need to understand this material
to understand the main ideas of this paper or to use the compiler or
the compiled code.
We are describing it here to illustrate how other \R\ programmers can
readily experiment with these tools to generate code in different
ways.

We'll use the Fibonacci sequence and the \Rfunc{fib} function example again,
as it illustrates a few different aspects of generating code.

Our \Rfunc{fib} function in \R\ expects an integer value and returns an
integer. The body consists of a single \Rkeyword{if--else} expression.
This contains a condition to test and two blocks of code, one of which
will be
evaluated depending on the outcome of that condition.
To map this code to \llvm\ concepts, we need to create different
instruction blocks, each of which contains one or more instructions.
When we call the routine, the evaluation starts in the first
instruction block and
executes each of its instructions sequentially. The end of each
instruction block
has a terminator which identifies the next block to which to jump, or
returns from the routine. Jumping between blocks allows us to
implement conditional branching, loops, etc.

%
\begin{figure}
{\fontsize{7.8}{9.8}\selectfont{\begin{verbatim}
; ModuleID = 'fib'

define i32 @fib(i32 %n) {
entry:
  %0 = icmp slt i32 %n, 2
  br i1 %0, label %"body.n < 2L", label %body.last

"body.n < 2L":                     ; preds = %entry
  ret i32 %n

body.last:                         ; preds = %entry
  %"n - 1L" = sub i32 %n, 1
  %1 = call i32 @fib(i32 %"n - 1L")
  %"n - 2L" = sub i32 %n, 2
  %2 = call i32 @fib(i32 %"n - 2L")
  %"fib(n - 1L) + fib(n - 2L)" = add i32 %1, %2
  ret i32 %"fib(n - 1L) + fib(n - 2L)"
}
\end{verbatim}}}%
\caption{Intermediate Representation for the compiled
\Rfunc{fib} routine. We create the \Cclass{Function} and
\Cclass{Block} objects and create and insert \llvm\ instruction
objects corresponding to the expressions and sub-expressions in the
\R\ function. The result is this low-level description in
intermediate form which \llvm\ can optimize and compile to native
code for different targets, for example, a CPU or GPU.
The different blocks have a label (e.g., \texttt{body.last})
and correspond to different parts of an \Rkeyword{if} statement or
possibly parts of a loop, generally.}\label{fig:IRCode}
\end{figure}

For our \Rfunc{fib} function, we start with an entry block that might
create any local variables for the computations. In our function,
this block simply contains code to evaluate the condition \Rexpr{n < 2}
and, depending on the value of this test, the instruction to branch to
one of two
other blocks corresponding to the expressions in the \Rkeyword{if} and
the \Rkeyword{else} parts. In the \Rkeyword{if} block (i.e., $n$ is
less than $2$), we add a single instruction to return the value of the
variable \Rvar{n}. In the block corresponding to the \Rkeyword{else}
part, we add several low-level instructions. We start by computing
\Rexpr{n - 1L} and then call \Rfunc{fib} with that value and store the
result in a local variable. Then we calculate \Rexpr{n - 2L}, call
\Rfunc{fib} and store that result. Then we add these two local
intermediate results and store the result. Finally we return that
result. Figure~\ref{fig:IRCode} shows the code in what is called
Intermediate Representation (IR) form that \llvm\ uses. This illustrates
the low-level computations.

While the code for this function is reasonably simple, there are many
details involved in generating the native code, such as defining the
routine and its parameters, creating the instruction blocks, loading
and storing
values, and creating instructions to perform subtraction, call the
\Cfunc{fib} function and return a value. The \llvm\ \Cpp\ API
(Application Programming Interface) provides numerous classes and
methods that allow us to create instances of these conceptual items
such as \Cclass{Function}s, \Cclass{Block}s, many different types of
instructions and so on. The \OmgPkg{Rllvm} package provides an \R\
interface to these \Cpp\ classes and methods and allows us to create
and manipulate these objects directly within \R.
For example, the following code shows how we can define
the function, the entry instruction block and generate the call \Rexpr
{fib(n - 1)}:\vspace*{7pt}
\begin{verbatim}
mod = Module()
f = Function("fib", Int32Type,
  list(n = Int32Type), mod)
start = Block(f)
ir = IRBuilder(start)
parms = getParameters(f)
n.minus.1 = binOp(ir, Sub, parms$n,
  createConstant(ir, 1L))
createCall(ir, f, n.minus.1)
\end{verbatim}\vspace*{7pt}
%

We don't want to write this code manually ourselves in \R, although
\OmgPkg{Rllvm} enables us to do so. Instead, we want to programmatically
transform the \R\ code in the \Rfunc{fib} function to create the
\llvm\ objects. The \mbox{\OmgPkg{RLLVMCompile}} package does this. Since
\R\ functions are regular \R\ objects which we can query and manipulate
directly in \R,
we can traverse the expressions in the body of a function, analyze
each one and perform a simple-minded translation from \R\ concepts to
\llvm\
concepts. This is the basic way the \Rfunc{compileFunction} generates
the code, using customizable handler functions for the different types
of expressions. These recognize calls to functions, accessing global
variables, arithmetic operations, if statements, loops and so on.
They use the functions in \OmgPkg{Rllvm} to create the corresponding
\llvm\ objects and instructions.

Once the compiler has finished defining the instructions for our
routine, \llvm\ has a description of what we want to do in the form of these
blocks and instructions. This description is in this intermediate
representation (IR). We can look at this \dquote{code} and it will look
something similar to that shown in Figure~\ref{fig:IRCode}. The IR code
shows the somewhat low-level details of the blocks and instructions as
we described above. We see the three blocks labeled \verb+entry+,
\verb+body.n < 2L+ and \verb+body.last+. Again, it is not important to
understand these details to be able to use the compiled routine
translated from the \R\ function. I show it here to illustrate the
different steps in the compilation process and to indicate that an
\R\ programmer can chose to change any of these steps.

Next, we instruct \llvm\ to verify and optimize the code. At this point,
we can call the new routine via the \Rfunc{.llvm} function in
\OmgPkg{Rllvm} which corresponds to a method in the \llvm\ API. The
first time the code is used, \llvm\ generates the native code from
the IR form.

\section{Contrasts with Related Research}\label{sec:OtherResearch}

There have been several projects exploring how to improve the
performance of \R\ code. We discuss some of these in this section.

\textit{Byte-Compiler}: One of the most visible projects is the byte-code
compiler developed by Luke Tierney \citep{bib:TierneyByteCompiler}.
This consists of an \R\ package
that provides the compiler and some support in the core \R\
interpreter to execute the resulting byte-compiled code. The compiler
maps the \R\ code to instructions in the same spirit as \llvm's
instructions and intermediate representation. These instructions are
at a higher-level than \llvm's and are more specific to~\R.

The typical speedup provided by the byte-compiler is a factor of
between $2$ and $5$, with much larger
speedups on some problems. This may not be sufficient to obviate the
need for writing code in \C/\Cpp. We probably need to see a factor of
more than $10$ and closer to $100$ for common tasks.

The byte-code compiler is written in \R\ and so others can adapt and
extend it. However, the details of how the resulting byte-code it generates
is evaluated is tightly embedded in the \C-language
implementation of the \R\ engine. This means that if one wants to
change the byte-code interpreter, one has to modify the \R\
interpreter itself. While one can do this with a private version of \R,
one cannot make these changes available to others without them also
compiling a modified version of \R. In other words, the byte-code
interpreter is not extensible at run-time or by regular \R\ users.
Furthermore, the \R\ core development team does not always greet
suggested enhancements and patches with enthusiasm. Therefore, this
approach tends to be the work of one person and so has limited
resources.

\textit{Ra JIT Compiler}: The Ra extension to $R$ and the associated
\Rpkg{jit} package is another approach to using JIT (Just-in-Time)
compilation in \R. This focuses on compiling loops and arithmetic
expressions in loops. Like the byte-code compiler above, Ra requires
almost no change to existing \R\ code by \R\ users---only the call
to the function \Rfunc{jit} before evaluating the code. The
performance gain on some problems can be apparently as high as a
factor of 27. (See
\url{http://www.milbo.users.sonic.net/ra/times9.html}.)
Unfortunately, this is no longer maintained on CRAN, the primary
central repository of \R\ packages. This approach suffers from the
fact that it requires a modified version of the \R\ interpreter,
again compiled from the \C-level source. This places a burden on the
author of Ra to continually update Ra as \R\ itself changes. Also, it
requires users trust Ra and take the time to build the relevant binary
installations of Ra.

As we mentioned previously, important motivating goals in our work are
to avoid
modifying \R\ itself, to allow other people to build on and adapt our
tools, and to directly leverage the ongoing work of
domain experts in compiler technology by integrating their tools to
perform the compilation. Our approach differs from both the
byte-compiler and Ra in these respects.

\textit{R on the Java Virtual Machine}: There are several projects
working on developing an implementation of \R\ using the \Java\
programming language and virtual machine. One is FastR
(\url{https://github.com/allr/fastr}), which is being developed by
a collaboration between researchers at Purdue, Oracle and INRIA.
Another is Renjin (\url{https://code.google.com/p/renjin/}).
Having \R\ run on the \Java\ virtual machine offers several
benefits. There are many interesting large-data projects implemented
in \Java, for  example, Apache's Hadoop and Mahout. Integrating \R\
code and
such projects and their functionality would be much tighter and
effective if they are all on the same platform and share the same
computational engine. Importantly, \R\ would benefit from passively
acquiring features in \Java\ and its libraries, for example, security,
threads. One very interesting development is that researchers in
Oracle, collaborating with the developers of FastR, are creating a
tool Graal (\url{http://openjdk.java.net/projects/graal/}) for
compiling the code ordinarily interpreted by a high-level virtual machine,
for example, FastR. This could yield the
performance gains we seek, but by passively leveraging the general
work of others merely by using widespread technologies. This contrasts
with the ongoing development of \R\ by a relatively small community
and having to actively and manually import new technologies, features
and ideas from other languages, systems and communities.

\textit{Translating \R\ code to \C}: Another approach is to translate \R
\ code to \C/\Cpp\ code. This is
attractive, as it would give us similar speedup as we can get with
\llvm, potentially produces human-readable code, and allows us to
leverage the standard tools for these languages such as compilers,
linkers and, importantly, debuggers. We can also potentially reuse
(some of) the generated code outside of $R$. Simon Urbanek's \Rpkg{r2c}
package \citep{bib:r2c} is an example of exploring translation of \R\
code to \C\ code.

The \Rpkg{Rcpp} \citep{bib:Rcpp} (and \Rpkg{inline}) package are widely used
in \R\ to improve performance. The packages provide a way to include
high-level \Cpp\ code within \R\ code and to compile and call it
within the \R\ code. The \Cpp\ code uses an $R$-like syntax to make
it relatively easier to write the \Cpp\ code. This has been a
valuable addition to \R\ to obtain performant code. The approaches that
compile \R\ code directly are preferable if they can get the same performance.
The first reason is because the programmer does not have to program in
\Cpp. It is also harder for other programmers to read the code and
understand what it does. The second reason is because the \Cpp\ code is
essentially opaque to any \R\ code analysis or compiler. If we do
manage to generally compile \R\ code effectively or implement an automated
parallel computing strategy for \R\ code, the \Cpp\ code cannot
easily be part of this. For example, if we can map the \R\ code to run
as a GPU kernel on many cores, we cannot easily combine the \R\ and
\Cpp\ code to take advantage of these cores.


\textit{Parallel Speedup}: There are several interesting proj\-ects that
have aimed at improving the performance of \R\ exclusively by running
the code in parallel. This is very important and in some sense
orthogonal to compilation of \R\ code. If we speed up the
computations on a single CPU, that speedup will benefit running code
on each CPU. However, we also want to compile \R\ code to take
advantage of multiple CPU/GPUs.\vadjust{\goodbreak} We hope to be able to integrate ideas
from these projects into our compilation strategies. Unfortunately,
some of them are no longer active projects, for example, pR and \Rpkg{taskPR}.
This illustrates one aspect we have observed in the \R\
community. Some researchers implement some ideas in \R, sometimes as
a PhD thesis, and then move on to other projects. One of the terrific
aspects of \R\ is the ongoing commitment to support the \R\
community. This is probably a very significant reason for \R's
widespread use and an important consideration when developing new
environments and languages. It is one of the forces motivating our
continued work within $R$, even if developing a new system would be
intellectually more stimulating.

A very important aspect of all this work is to recognize that there
are many positive ways to make \R\ faster and more efficient. While
one of these approaches \textit{may} dominate others in the future, it
is very important that we should pursue comparative approaches and
continue to motivate each other's work. There is much to be learned
from these different approaches that will improve the others.

\section{Future Work}\label{sec:FutureWork}

Compiling (subsets of) \R\ code and other Domain Specific Languages
(DSLs) within \R\ using
\llvm\ is a promising approach that is certainly worth vigorously
pursuing in the near term. The work is currently in its infancy---we
started it in the summer of 2010, but have only recently returned to
it after an almost three-year hiatus due to other projects (ours and
other people's). However, the foundations of
many of the important components are in place, that is, the \OmgPkg{Rllvm}
package, and the basics of the extensible and adaptable compiler
mechanism in \OmgPkg{RLLVMCompile} should allow us and others to make
relatively quick progress, programming almost entirely in \R\ to
develop compilation strategies. However, there are many other tasks to
do to make
these transparent and reliable, and many related projects that will
make them more
powerful and convenient.

One of the immediate tasks we will undertake is to program some rich
examples explicitly in \R\ code. We are implementing \R\ code versions
of recursive partitioning trees, random forests and boosting.
We also plan to explore compiling code for the Expectation
Maximization (EM) algorithm and particle filters to run on GPUs. The
aim is to share these sample \R\ projects with the other researchers
investigating different compilation strategies for \R\ so that we
compare approaches on substantive and real tasks we want to program in pure
\R\ code.\vadjust{\goodbreak}

We plan to add some of the functionality available in \llvm\ that
does not yet have bindings in \OmgPkg{Rllvm}. This includes topics
such as different optimization passes and adding meta data to the
instructions. We have also developed the initial infrastructure to compile
\R\ code as kernel routines that can be used on GPUs, that is, PTX
(Parallel Thread Execution) code. Being able to generate kernel
functions from \R\ code, along with the existing \R-CUDA bindings to
manage memory and launch kernels from the host device, allows us to
program GPUs directly within high-level \R\ code. This contrasts
with the low-level C code developed for existing \R\ packages that
target GPUs, for example, \Rpkg{gputools} \citep{bib:gputools} and
\Rpkg{rgpu} \citep{bib:rgpu}.

We will also be exploring different approaches to compiling the \R\
code to run in parallel and distributed settings. We think that being
able to use information about the distribution of the data to
generate/compile the code will be important so that we can minimize
the movement of data and keep the CPUs/GPUs busy on the actual
computations rather than transferring the inputs and outputs to and
from the computations.

Being able to write \R\ code that directly calls \C\ routines is
very powerful. As we saw in relation to the \Cfunc{fgets} routine in
Section~\ref{eg:SamplingCSV}, we need to specify the signature for the
routines we want to call. It is preferable to be able to
programmatically identify these signatures rather than require \R\
programmers to explicitly specify them. The \OmgPkg{RCIndex}
package is an \R\ interface to \lib{clang} \citep{bib:libclang}, the
parsing facilities for the clang compiler. This already allows us to
read \C\ and \Cpp\ code in \R\ and to identify the different
elements it contains. This allows to not only determine the
signatures of routines, but also discover different data structures,
enumerated constants, etc. We can also go further and understand more
about how the routines manipulate their arguments and whether they
perform the memory management or leave it to the caller.

As we saw in each of our examples, information about the types of each
parameter and local variable is a necessity to being able to compile
using \llvm. Currently, the \R\ programmer must specify this
information not only for the function she is compiling, but also for
all of the
functions it calls. Again, we want to make this transparent, or at
least only require the \R\ programmer to specify this information when
there is
ambiguity. To this end, we are working on a type inference package
for \R. This starts with a known set of fundamental functions and
their signatures. From this, we can determine the signatures of many
higher level calls. As always, we cannot deal with many features of
the language such as nonstandard evaluation, but we most likely can
get much of the type information we need programmatically. Since \R's
types are so flexible with different return types based on not only
the types of the inputs, but also the content of the inputs, we need a
flexible way to specify types. Perhaps the existing \BiocPkg{TypeInfo}
package \citep{bib:TypeInfo} or \Rpkg{lambda.r} package will help here.
To analyze code for type
information and for variable dependencies, we will build upon the
\OmgPkg{CodeDepends} \citep{bib:CodeDepends} and
\Rpkg{codetools} \citep{bib:codetools} packages.

While these are some of the related activities we envisage working on,
we also encourage others to collaborate with us or work independently
using \llvm\ and optionally \OmgPkg{Rllvm} and \OmgPkg{RLLVMCompile}
so that our community ends up with better tools.

\section{Conclusion}

We have described one approach to making some parts of the \R\
language fast. We leverage the compiler toolkit infrastructure \llvm\
to generate native code. This allows us to incorporate technical
knowledge from another community, both now and in the future. We can
generate code for CPUs, GPUs and other targets. We can dynamically
specialize \R\ functions to different computational approaches, data
representations and sources, and contextual knowledge, giving us a new
and very flexible approach to thinking about high-level computing.

We are developing a simple but extensible and customizable compiler
in \R\ that can translate \R\ code to native code.
Not only does this make the code run fast, but it also allows us to
compute in quite different ways than when we interpret the \R\ code
in the usual way. We can even outperform some of \R's own native code.

In no way should this work be considered a general compiler for all of
the \R\ language. There are many aspects of the \R\ language we
have not yet dealt with or considered. Vectorized subsetting,
recycling, lazy evaluation and nonstandard evaluation are examples. We,
or others, can add facilities to the compiler to support these when
they make sense and are feasible.

The initial results from this simple approach are very encouraging.
An important implication of this and other efforts to make \R\ code
efficient is that we can be benefit from writing high-level code that
describes what to compute, not how. We then use smart interpreters or
compilers to generate efficient code, simultaneously freeing \R\
programmers to concentrate on their tasks and leveraging domain
expertise for executing the code. We hope others will be able to use
these basic building blocks to improve matters and also to explore
quite different approaches and new languages within the \R\ environment.

\section*{Acknowledgments}

Vincent Buffalo made valuable contributions to designing and
developing the \textbf{\textit{RLLVMCompile}} package in the initial work.
Vincent Carey has provided important ideas, insights, advice and
motivation and I~am very grateful to him for organizing this
collection of papers and the session at the 2012 Joint Statistical
Meetings. Also, I appreciate the very useful comments on the initial
draft of this paper by the three reviewers and also John Chambers.

\section*{Supplementary Material}

The code for the examples in this paper, along with the timing results
and their meta-data, are available from
\texttt{\href{https://github.com/duncantl/RllvmTimings}{https://github.com/duncantl/}\break
\href{https://github.com/duncantl/RllvmTimings}{RllvmTimings}} as a git repository.
The versions of the \textbf{\textit{Rllvm}} and \mbox{\textbf{\textit{RLLVMCompile}}}
packages involved in the timings can also be retrieved from their
respective git repositories. The specific code used is associated with
the git tag \textit{StatSciPaper}.


%

\end{document}